\documentclass[referee]{raa}            

\usepackage{graphicx,times}             
\usepackage{natbib}
\usepackage{amssymb,amsmath}
\bibpunct{(}{)}{;}{a}{}{,}

\usepackage[a4paper=true,pdftex=true,pagebackref=true]{hyperref}
\hypersetup{colorlinks = true, linkcolor = green, anchorcolor = red, citecolor = blue, filecolor = red, pagecolor = red, urlcolor = red}

\begin{document}
	
	\title{DES Map Shows a Smoother Distribution of Matter than Expected:
		A Possible Explanation
	}
	
	\volnopage{Vol.0 (20xx) No.0, 000--000}      
	\setcounter{page}{1}          
	
	\author{E.Oks
		\inst{}
	}

\institute{Physics Department, 380 Duncan Drive, Auburn University, Auburn, AL 36849, USA; {\it oksevgu@auburn.edu}\\
	\vs\no
	{\small Received~~20xx month day; accepted~~20xx~~month day}}

	\abstract{ The largest and most detailed map of the distribution of dark matter in the Universe has been recently created by the DES team. The distribution was found to be slightly (be few percent) smoother, less clumpy than predicted by the general relativity. This result was considered as a hint of some new physical laws. In the present paper we offer a relatively simple model that could explain the above result \textit{without resorting to any new physical laws}. The model deals with the dynamics of a system consisting of a large number of gravitating neutral particles, whose mass is equal to the mass of hydrogen atoms. The central point of the model is a partial inhibition of the gravitation for a relatively small subsystem of the entire system. It would be sufficient for this subsystem to constitute just about few percent of the total ensemble of the particles for explaining the few percent more smooth distribution of dark matter (observed by the DES team) compared to the prediction of the general relativity. The most viable candidate for the dark matter particles in this model is the second flavor of hydrogen atoms (SFHA) that has only S-states and therefore does not couple to the electric dipole radiation or even to higher multipole radiation, so that the SFHA is practically dark. The SFHA has the experimental confirmation from atomic experiments, it does not go beyond the Standard Model, it is based on the standard quantum mechanics, and it explains the puzzling astrophysical observations of the redshifted line 21 cm from the early Universe. Thus, our model explaining the DES result of a little too smooth distribution of dark matter \textit{without resorting to any new physical laws} seems to be self-consistent.
	}
	
	\authorrunning{E.Oks}        
	\titlerunning{DES Map Shows a Smoother Distribution of Matter than Expected }  
	
	\maketitle
	
	%
	%
	\section{Introduction}           
	\label{sect:intro}
	
	The largest and most detailed map of the distribution of dark matter in the Universe has been recently created by the DES team (\citealt{Chang+etal+2018},\citealt{Jeffrey+etal+2021}). The distribution was found to be slightly (be few percent) smoother, less clumpy than predicted by the general relativity (\citealt{Jeffrey+etal+2021}). This result was considered as a hint of some new physical laws.
	In the present paper we offer a relatively simple model that could explain the above result \textit{without resorting to any new physical laws}. The model deals with the dynamics of a system consisting of a large number of gravitating neutral particles, whose mass is equal to the mass of hydrogen atoms. The central point of the model is a partial inhibition of the gravitation for a relatively small part of the system.

	\section{The model}
	\label{sect:Obs}
	
    We consider a system of a large number of gravitating neutral particles of the mass M equal to the mass of hydrogen atoms. (The nature of the particles will be specified later on.) At any instant of time, the system has a subsystem of relatively isolated pairs of particles, i.e., pairs where the separation within the pair is much smaller than their distance to other particles. The subsystem is open. This means that after some time, some pairs would not qualify any more as the subsystem members (because they can no more be considered as relatively isolated), while some other pairs could become relatively isolated and qualify as new members of the subsystem. Here the word “subsystem” means a subset of particle within the ensemble – the subset of pairs (not located in one particular volume) that are relatively isolated.
	The pairs loose the energy by the gravitational radiation and the separation within the pair decreases. This is similar to the classical description of a usual hydrogenic atom or ion: it emits electromagnetic radiation and the separation between the electron and the nucleus decreases. While classically the latter process would lead to the fall of the electron into the nucleus, in the quantum description there arises the average minimum separation $R_{min}$ (the Bohr radius in the case of hydrogen atoms), at which the “fall” of the electron into the nucleus stops:
	\begin{equation}	
		R_{min}= \hbar^2/(\mu\alpha).
		\label{Rmin}
	\end{equation}
     In Eq. (\ref{Rmin}), $\mu$ is the reduced mass of the pair and $\alpha$ is the coupling coefficient in the corresponding potential energy $V$:
     \begin{equation}
      V = –\alpha/R.
      \label{potV}
    \end{equation}
    Similarly, for the pairs of gravitating particles, there is the average minimum separation, at which the gravitational radiation stops and there is no further decrease of the separation within the pair. In this situation, one has 
    \begin{equation}
    	\mu=M/2,   \;  \alpha=GM^2, 
    	\label{MuAlpha}
    \end{equation}
	(where $G$ is the gravitational constant), so that 
	\begin{equation}
		R_{min}= 2\hbar^2/(GM^3).
		\label{Rmin=}
	\end{equation}
	
	On substituting in Eq. (\ref{Rmin=}) the mass M equal to the mass of hydrogen atoms, we get $R_{min}$ = 2.3 megaparsecs. This is practically the same as the average observed separation between galaxies.
	So, at the separation within the pair of the gravitation particles $R\sim R_{min}$, their further approach to each other stops. This is equivalent to a partial inhibition of the classical gravitation. 
	In other words, if the minimum non-zero separation $R_{min}$ did not exist, then the average separation within any such pair of two gravitating hydrogen atoms would become smaller and smaller without any lower limit (corresponding to the uninhibited “clumping” of dark matter). However, due to the existence of the minimum non-zero separation $R_{min}$, the further “clumping” becomes inhibited for such pairs. In terms of the total ensemble of hydrogen atoms, the further “clumping” becomes inhibited only for a relatively small part of the ensemble: for the subsystem of such pairs. We note that because the subsystem of such pairs of hydrogen atoms is relatively small, this effect manifests only in some additional smoothness of the dark matter distribution, but it does not manifest in the rotation curves of galaxies.
	Now let us address the following question: how much time does it take for the pair of gravitating hydrogen atoms to reach the state of the non-zero minimal separation. In Appendix A we show that the corresponding time $T$ scales with the initial angular momentum $L_0$ of the pair as $L_0^3$. Consequently, for sufficiently small $L_0$, the time $T$ can be much smaller than any benchmark. For instance, it can be not only much smaller than the age of the Universe, but also much smaller than the duration of any specific stage of the Universe evolution.
	Next, we estimate the percentage of such pairs of hydrogen atoms in the total ensemble of hydrogen atoms. According to quantum mechanics, for a given principal quantum number $n$, describing the gravitating pair of hydrogen atoms, the angular momentum $L_0$ can take the following $n$ values: $0, \hbar, 2\hbar, …, (n – 1)\hbar$, where $\hbar$ is the Planck constant. Let us assume that only for $L_0 = 0$, the time $T$ is smaller than the corresponding benchmark time, characterizing the Universe or its stage of the evolution. In the subsystem of pairs, each pair can be initially in the state of some particular principal quantum number $n \leqslant n_{max}$, where the value of $n_{max} \gg 1 $ will be estimated in the next paragraph. Then the total number $K$ of possible values of the initial values $L_0$ of the angular momentum is the following
	\begin{equation}
		K=\sum_{n=1}^{n_{max}}n=n_{max}(n_{max}+1)/2\approx n_{max}^2/2,
		\label{Ksum}
	\end{equation}
	where in the utmost right side we used the inequality $n_{max} \gg 1 $. The number of pairs of $L_0 = 0$ is 
	\begin{equation}
		k=\sum_{n=1}^{n_{max}}1=n_{max}.
		\label{k}
	\end{equation}
	Thus, the share of pairs, having enough time to reach the state of the non-zero minimal separation (the ground state, corresponding to the inhibition of the gravitational interaction), is
	\begin{equation}
	    k/K\approx 2/n_{max}
	    \label{kK}
	\end{equation}
The value of $n_{max}$ can be estimated as follows. According to quantum mechanics, the average size $R$ of a pair of hydrogen atoms in the state of the principal quantum number $n$ is
   \begin{equation}
   	R\sim n^2R_{min},
   	\label{Rapp}
   \end{equation}
where $R_{min}$ is given by Eq. (\ref{Rmin=}). The maximum possible value of $R$ should be smaller or of the order of magnitude of the radius of the Universe $R_U$. Consequently, we get
\begin{equation}
	n_{max}\lesssim(R_U/R_{min})^{1/2}
	\label{nmaxLess}
\end{equation}	
Since $R_{min} \sim 2.3$ megaparsecs and $R_U \sim 1.4$x$10^4$ megaparsecs, Eq. (\ref{nmaxLess}) yields $n_{max} \lesssim 80$. Then, according to Eq. (\ref{kK}), the share of pairs, having enough time to reach the ground state, (corresponding to the inhibition of the gravitational interaction) is 
\begin{equation}
	k/K\gtrsim 1/40=2.5\%
\end{equation}
This estimate of the percentage of the pairs, exhibiting the inhibition of the gravitational interaction and thus the inhibition of the unlimited “clumping”, agrees with (and could explain) the few percent more smooth, less clumpy distribution of dark matter (observed by the DES team) compared to the prediction of the general relativity.
\par
\begin{figure}[h]
	\begin{minipage}[t]{0.495\linewidth}
		\centering
		\includegraphics[width=60mm]{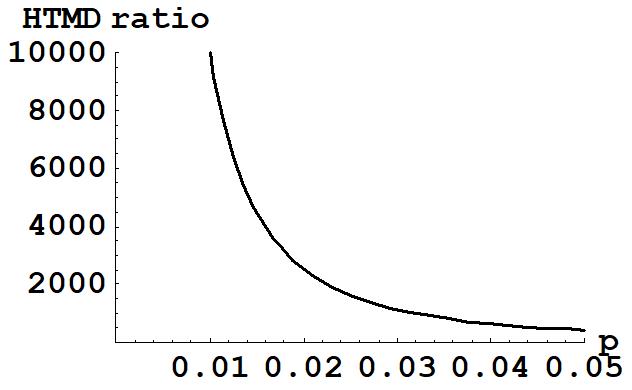}
	\end{minipage}%
	\begin{minipage}[t]{0.495\textwidth}
		\centering
		\includegraphics[width=60mm]{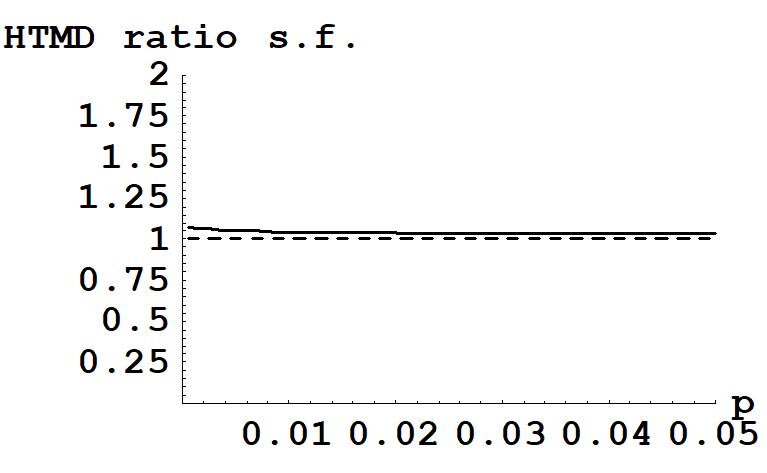}
	\end{minipage}%
	\caption{{\small  Ratio of the theoretical and experimental results for the High-energy Tail of the linear Momentum Distribution (HTMD). Left panel: for the ground state of the usual hydrogen atoms. Right panel: for the ground state of the second flavor (s.f.) of hydrogen atoms shown by the solid line; the dashed line is the horizontal line at HTMD = 1 shown for visualizing how close the solid line is to unity. The linear momentum p is in units of mc, where m is the electron mass and c is the speed of light.}}
	\label{fig12}
\end{figure}

 Now let us address the question whether dark matter particles can have the mass of hydrogen atoms. There is a definitive proof from atomic experiments that there are two kinds – or two flavors – of hydrogen atoms: the usual ones and the second flavor (\citealt{Oks+2001}). Only the existence of this alternative kind of hydrogen atoms eliminated a huge (many orders of magnitude) discrepancy between the experimental and theoretical results for the High energy Tail of the linear Momentum Distribution (HTMD) in the ground state of hydrogen atoms (\citealt{Oks+2001}) – see Fig. \ref{fig12}

The second flavor of hydrogen atoms (SFHA) has only states of zero angular momentum (the S-states) both in the discrete and the continuous spectra (\citealt{Oks+2001},\citealt{Oks+2020a},\citealt{Oks+2020b}). Therefore, due to the selection rules the SFHA does not have states that can be coupled by the electric dipole radiation, as noted in those papers.    
Moreover, the SFHA cannot be coupled by electric quadrupole or a higher multipole radiation either. (This fact is new: it was not mentioned in the above-referenced papers.) This is because the quadrupole, octupole, and all higher multipole terms, containing linear combinations of various powers of the radius-vector operator $\textbf{r}$ of the atomic electron, yield zeros in all orders of the perturbation theory – both diagonal and non-diagonal matrix elements of the operator $\textbf{r}$ are zeros for the S-states. Therefore, the SFHA is practically “dark”: it is dark in all ranges of the electromagnetic spectrum except for the 21 cm line, corresponding to the radiative transition between the hyperfine sublevels of the ground state.
In addition to the definitive proof from atomic experiments, there seems to be also another evidence in favor of the existence of the SFHA – from astrophysics. Their existence can explain recent puzzling astrophysical observations concerning the redshifted radio line 21 cm from the early Universe \citealt{Bowman+etal+2018}: the explanation presented in \cite{Oks+2020a} paper did not require resorting to hypothetical, never discovered subatomic particles – in distinction to the explanation from \cite{Barkana+2018} paper\footnote{We note that \cite{McGaugh+2018} examined the results by \cite{Bowman+etal+2018} and by \cite{Barkana+2018}, and came to the following important conclusion: the observations by \cite{Bowman+etal+2018} constitute an unambiguous proof that dark matter is baryonic, so that models introducing non-baryonic nature of dark matter have to be rejected.}.
   It should be emphasized that the SFHA did not result from changing physical laws. In fact, the description of the SFHA is based on the standard quantum mechanics: namely, on the employment of the so-called “singular” solution of the Dirac equation (\citealt{Oks+2001}).
The SFHA is the only one candidate for dark matter that has the following four features simultaneously: 1) it has the experimental confirmation, namely from the analysis of atomic experiments (\citealt{Oks+2001}); 2) it does not go beyond the Standard Model (and thus is favored by the Occam razor principle); 3) it is based on the standard quantum mechanics – without any change to the physical laws (and thus is favored by the Occam razor principle); 4) it explains the puzzling astrophysical observations by \cite{Bowman+etal+2018}.
Thus, the SFHA is a viable candidate for dark matter or at least a part of it. Therefore, the above model explaining the DES result of a little too smooth distribution of dark matter seems to be self-consistent.

	\section{Conclusions}
	\label{sect:conclusion}

We analyzed a system of a large number of gravitating neutral particles, whose mass is equal to the mass of hydrogen atoms, and focused at the subsystem of relatively isolated pairs of these particles. The pairs loose the energy by the gravitational radiation and the separation within the dowpair decreases. We showed that this process would stop as the separation of the particles within the pair would decrease to the minimum value of the order of few megaparsecs. This minimum value is practically the same as the average observed separation between galaxies.
The termination of the gravitational radiation and of the further decrease of the separation of the particles with the pairs is equivalent to a partial inhibition of the classical gravitation. Our estimate of the percentage of the pairs, exhibiting the inhibition of the gravitational interaction and thus the inhibition of the unlimited “clumping”, is $\gtrsim 2.5\%$. This agrees with the percentage observed by the DES team: the few percent more smooth, less clumpy distribution of dark matter compared to the prediction of the general relativity.
Dark matter particles having the mass of hydrogen atoms could be the second flavor of hydrogen atoms (SFHA) that has only S-states and therefore does not couple to the electric dipole radiation or even to higher multipole radiation, so that the SFHA is practically dark \footnote{\textbf{There is also a slight difference between the SFHA and usual hydrogen atoms in cross-sections of
	charge exchange with an incoming proton \citealt{Oks+2021}.}}. The 
SFHA has the experimental confirmation from atomic experiments, it does not go beyond the 
Standard Model, it is based on the standard quantum mechanics, and it explains the puzzling astrophysical observations by \cite{Bowman+etal+2018}. Thus, our model explaining the DES result of a little too smooth distribution of dark matter \textit{without resorting to any new physical laws} seems to be self-consistent. While the model is relatively simple – just to get the message across – we hope it would motivate further studies.

	\appendix                  
	\section{Time required to reach the state of the non-zero minimal separation}
	   For a pair of two hydrogen atoms revolving in elliptical orbits about their barycenter, the loss of the energy $E$ and of the angular momentum $L$ (both averaged of the period of the motion) per unit time due to the gravitational radiation is given by the following expressions (see, e.g., the textbook by \cite{Landau+Lifshitz+1975}):
	\begin{equation}
		d|E|/dt = 64G^4M^5(1 + 73e^2/24 + 37e^4/96)/[5c^5a^5(1 – e^2)^{7/2}], 
		\label{de/dt}
	\end{equation}
	\begin{equation}
		dL/dt = – 2^{11/2}G^{7/2}M^{9/2}(1 + 7e^2/4)/[5c^5a^{7/2}(1 – e^2)^2].	
		\label{dl/dt}
	\end{equation}
In Eqs. (\ref{de/dt}) and (\ref{dl/dt}), $G$ is the gravitational constant, $M$ is the hydrogen atom mass (to distinguish from which we denoted the angular momentum by $L$); $a$ and $e$ are the major semi axis and the eccentricity of the elliptical orbit, respectively:
	\begin{equation}
		a = GM^2/(2|E|), \;	e = [1 – 4|E|L^2/(G^2M^5)]^{1/2}.
		\label{a,e}
	\end{equation}
We are interested in elliptical orbits of a relatively large eccentricity, such as
	\begin{equation}
		(1 – e^2) = 4|E|L^2/(G^2M^5) \ll 1.
		\label{e2}
	\end{equation}
In this situation, Eqs. (\ref{de/dt}) and  (\ref{dl/dt}) simplify as follows:
	\begin{equation}
		d|E|/dt \approx 170G^4M^5/[3c^5a^5(1 – e^2)^{7/2}], 	
		\label{de/dt2}
	\end{equation}
	\begin{equation}
		dL/dt \approx – 2^{7/2}11G^{7/2}M^{9/2}/[5c^5a^{7/2}(1 – e^2)^2].
		\label{dl/dt2}
	\end{equation}
After dividing Eq. (\ref{de/dt2}) by Eq. (\ref{dl/dt2}) and substituting $a$ and $e$ from Eq. (\ref{a,e}), we get:
	\begin{equation}
		d|E|/dL \approx – 425G^2M^5/(264L^3).
		\label{de/dl}
	\end{equation}
By integrating Eq. (\ref{de/dl}), we obtain
	\begin{equation}
		|E| \approx |E_0| + (425/528)G^2M^5(1/L^2 – 1/L_0^2), 	
		\label{E}
	\end{equation}
where $E_0$ and $L_0$ are the initial values of the energy and angular momentum, respectively.
On substituting in Eq. (\ref{dl/dt2}) the expressions for a and e from Eq. (\ref{a,e}), as well as the expression for $|E|$ from Eq. (\ref{E}), we find
	\begin{equation}
		dL/dt \approx – [88G^4M^{15/2}|E_0|^{3/2}/(5c^5L^7)][L^2 + b(L_0^2 – L^2)]^{3/2},
		\label{dL/dt}
	\end{equation}
where
	\begin{equation*}
		b = 425G^2M^5/(528L_0^2|E_0|).	
	\end{equation*}
After integrating Eq. (\ref{dL/dt}) in the limits from $L_0$ to $0$ (with respect to $L$) and from $0$ to $T$ (with respect to $t$), we obtain:
	\begin{equation}
		T \approx c^5L_0^5(1 + 4b^{1/2} + 5b)/[88G^4M^{15/2}|E_0|^{3/2}].	
		\label{T1}
	\end{equation}
From Eq. (\ref{e2}) it follows that $b \gg 1$, so that $(1 + 4b^{1/2} + 5b) \approx 5b$ and Eq. (\ref{T1}) simplifies to the following final form:
	\begin{equation}
		T \approx 2125c^5L_0^3/(46464G^2M^{5/2}|E_0|^{5/2}).
		\label{T2}
	\end{equation}
From Eq. (\ref{T2}) it is seen that the time $T$, required for two gravitating hydrogen atoms to reach the state of the non-zero minimal separation (the ground state), scales with the initial angular momentum $L_0$ as $L_0^3$. Therefore, for sufficiently small $L_0$, the time $T$ can be much smaller than any benchmark: e.g., much smaller than the age of the Universe or even much smaller than the duration of any specific stage of the Universe evolution.

\label{lastpage}
	
\end{document}